\documentclass[twocolumn]{aastex63}

\usepackage{float,graphicx,amsmath,multirow}
\usepackage{color}
\usepackage{chemformula}

\usepackage{booktabs}

\usepackage{lineno}

\usepackage{xcolor}
\usepackage{xifthen}
\usepackage[normalem]{ulem}
\newcommand{\stkout}[1]{\ifmmode\text{\sout{\ensuremath{#1}}}\else\sout{#1}\fi}
\newcommand{\edited}[2]{\ifthenelse{\isempty{#1}}{\textbf{#2}}{\ifthenelse{\isempty{#2}}{\textcolor{gray}{\stkout{#1}}}{\textcolor{gray}{\stkout{#1}} \textcolor{red}{#2}}}}

\usepackage{xstring,xifthen}
\newcommand{\replacedecimal}[2]{\ifthenelse{\isin{.}{#1}}{\text{\StrBefore{#1}{.}}\ensuremath{\overset{#2}{.}}\text{\StrBehind{#1}{.}}}{#1\ensuremath{^{#2}}}}

\makeatletter
\newcommand*{\rom}[1]{\expandafter\@slowromancap\romannumeral #1@}
\makeatother
\newlength{\VSpaceBeforeTabBib}
\newlength{\VSpaceBeforeTabFoot}
\newcounter{posteriorsubfigure}
\newcounter{cachedposteriorsubfigure}

\begin{document}
\title{A Classifier Framework for ALMA Bandpass Calibration Anomalies}

\author[0000-0003-2760-2119]{Ci Xue}
\affiliation{National Radio Astronomy Observatory, Charlottesville, VA 22903, USA}
\affiliation{NSF–Simons AI Institute for Cosmic Origins, Austin, TX, 78712, USA}

\author[0000-0002-8472-836X]{Brian S. Mason}
\affiliation{National Radio Astronomy Observatory, Charlottesville, VA 22903, USA}

\author[0000-0003-0465-4944]{Gazi A. Rakib}
\affiliation{University of Utah, Salt Lake City, UT 84112, USA}

\author[0000-0002-7431-8543]{Tristan Ashton}
\affiliation{National Radio Astronomy Observatory, Charlottesville, VA 22903, USA}

\author[0000-0003-1169-2965]{Jeff M. Phillips}
\affiliation{University of Utah, Salt Lake City, UT 84112, USA}

\author[0000-0002-8086-8226]{Ignacio Toledo}
\affiliation{Joint ALMA Observatory, Santiago, Chile}

\author[0000-0002-8932-1219]{Ryan A. Loomis}
\affiliation{National Radio Astronomy Observatory, Charlottesville, VA 22903, USA}

\author[0000-0001-9163-0064]{Ilsang Yoon}
\affiliation{National Radio Astronomy Observatory, Charlottesville, VA 22903, USA}

\author[0000-0002-2202-608X]{John E. Hibbard}
\affiliation{National Radio Astronomy Observatory, Charlottesville, VA 22903, USA}

\author[0000-0003-2722-8841]{Omkar Bait}
\affiliation{National Radio Astronomy Observatory, Charlottesville, VA 22903, USA}
\affiliation{NSF–Simons AI Institute for Cosmic Origins, Austin, TX, 78712, USA}

\author[0000-0001-7089-7325]{Eric J. Murphy}
\affiliation{National Radio Astronomy Observatory, Charlottesville, VA 22903, USA}

\correspondingauthor{Ci Xue}
\email{cxue@nrao.edu}

\begin{abstract}
Bandpass calibration corrects instrumental responses across observing frequencies in interferometric radio observations, so the measured calibration solutions inherently contain spectral variation that varies with instrumental configurations and noise conditions. However, some signals contain anomalies, which if not caught can lead to poorly constructed science products. Observatory pipelines often use scalar metrics with individual, linear decision boundaries. Constructing a simple parametric model can be effective for observations with a relatively limited parameter space, but it becomes progressively less accurate as the parameter space expands. To address this challenge, we design a machine learning framework that treats anomaly identification as a supervised classification task and learns complex, non-linear decision boundaries across multiple features optimized for predictive performance. In particular, we use an XGBoost model that sits on top of several features and specialized scan statistic models to classify anomalies in amplitude solutions from ALMA bandpass calibrations. Trained on an extensive, quality-controlled ALMA dataset, this classifier framework leverages features derived from kernel-based scan statistics to characterize deviating spectral intervals, while remaining extensible to additional feature sets. This method substantially reduces misclassifications in bandpass anomaly detection compared to the current pipeline heuristics. By improving flagging decisions and calibration quality, this classifier framework provides a scalable supplement for observatory operations and is generalizable to next-generation radio facilities.
\end{abstract}

\keywords{Calibration, Radio interferometers, Radio observatories, Astronomy data analysis}

\section{Introduction \label{sec:intro}}
Processing data from general-purpose interferometric radio observatories is challenging, with one key challenge being distinguishing observations that are conducted under expected system performance, from observations that require corrective intervention. The Atacama Large Millimeter/sub-millimeter Array (ALMA) pipeline \citep{2023PASP..135g4501H} represents a revolutionary step toward automated processing and data delivery, but often requires human inspection to ultimately decide if the data can be delivered automatically or if corrective action must be taken. For the early years of ALMA operations through the end of 2023, data reducers inspected every reduction to identify calibration issues. More recently, ALMA introduced a quantitative Quality Assessment (QA) process that facilitates partial automation of pipeline processing and data delivery \citep{GarciaYus_2020, Nakos_2020}. Nevertheless, identifying calibration issues that compromise the scientific integrity of the data continues to rely either on human inspection or on expert-crafted heuristics with manually tuned thresholds, which may not generalize across observations.

One important step in the calibration process is bandpass calibration, which normalizes the instrumental and atmospheric transmission across frequencies \citep{remijan_2019_4511522, 2022ApJ...938..151E, 2025A&A...703A.196A}. Observations of bright, spectrally flat bandpass calibrators are used to measure the antenna responses as a function of observing frequency, referred to as calibration solutions. Each bandpass solution represents complex responses, in amplitude and phase, of frequency channels for each spectral window (SPW), antenna, and polarization. Outlier features in bandpass calibration solutions often show as artificial passband shape distortions, which can propagate into the downstream calibration process and introduce artifacts into the science product \citep[e.g.][]{2021NatAs...5..631V}.

As in the case of ALMA observations, example outlier features in bandpass solutions include large variability over spectral channels, spikes caused by instrumental leakage from the water vapor radiometer (WVR), and systematics introduced by the correlator itself. While some of these effects can be mitigated through standard calibration procedures, such as WVR corrections, others require identifying and flagging anomalous solutions and excluding them from the calibration process. One such effect is ``platforming'' or ``subbanding'', arising from occasional errors in the way that the correlator stitches together subsections of a given SPW \citep{cortes_2022_7764458}. This gives rise to discrete offsets in amplitude between adjacent subbands. Platforming anomalies are particularly common in wide-bandwidth correlators, and are therefore expected to remain important in next-generation radio facilities. The ATAC correlator at the heart of the upcoming ALMA Wideband Sensitivity Upgrade (WSU) \citep{2024pcaw.confE..12B, 2024stt..conf....1D} will implement a similar subband stitching procedure and may also have such  platforming anomalies, though on a wider scale.

Various statistical methods have been developed to identify outliers from the background distribution (including astronomical signals, measurement interference, and anomalies relative to the baseline or noise) and to quantify detections using statistical metrics \citep[e.g.][]{2010MNRAS.405..155O, 2010MNRAS.406L..60N, 6846297, 2019ApJ...871..112X, 2022RemS...14.4672P}. A conventional approach for setting a decision threshold is to obtain an empirical distribution of metric values from well-performing data and compare observed metrics against this empirical baseline. Such fixed-threshold hypothesis tests require the validation of the empirical baseline being representative. While tuning independent decision thresholds can be effective in a limited parameter space, it becomes impractical in a complex, sparsely sampled parameter space. As the number of input parameters and the possible combinations increase, the number of thresholds needed to be determined and validated also increases, because each parameter combination may have its own empirical baseline distribution and corresponding decision threshold. Meanwhile, as the parameter space expands, observations become sparse within each parameter combination, leaving too few samples to obtain representative empirical distributions. In practice, astronomical facilities with a complex parameter space, such as ALMA with varying antenna configurations and noise conditions, often apply a global threshold to each metric, which results in context-unaware decisions. Consequently, a given metric value can be normal for one input parameter combination but an outlier for another.

Machine learning (ML) methods, in contrast, excel at learning multivariate decision boundaries. The ALMA Science Archive contains over a decade of expert-curated, high-quality data that can serve to train a classifier. By using labels based on data reducers' inspection of these data, ML models can learn which metrics or features provide the best trade-off between the prediction purity and completeness by cross validation. One advantage of ML classification approaches over empirical statistical approaches is that ML reduces dependence on potentially fragile statistical assumptions and makes decisions directly by optimizing predictive performance.

In this work, we present a classifier framework for detecting anomalies in bandpass calibration solutions, with the current implementation focusing on the amplitude portion of complex solutions. In Section~\ref{sec:data}, we describe the dataset of bandpass solutions from selected ALMA observations. Section~\ref{sec:meth} presents the classifier framework, including the labeling strategy for both training and evaluation and the selection of input features derived from the raw solution data.

\section{Data \label{sec:data}}

\subsection{Bandpass Calibration Data}

Our models are trained on bandpass calibration data from selected public ALMA Principal Investigator (PI) science observations delivered during Cycle 9 (October 2022 -- September 2023). Until the middle of Cycle 10, all ALMA data were systematically reviewed by ALMA staff prior to delivery to PIs. To prepare for the beginning of automatic deliveries in Cycle 10, the ALMA Cycle 9 data were subject to additional, retrospective review. These Cycle 9 datasets--- and in particular the so-called ``False Negatives'' which were assessed to contain an actionable issue but that would have been auto-delivered with the QA assessment in place at the time--- are the training data for this work. This `false negative' dataset is preferentially enriched in features of interest for anomaly detection, making it well-suited for improving automated decision support. Among the anomalies present in this corpus, bandpass ``subbanding''  (discontinuities in the bandpass solution within a spectral window) is of particular relevance. This subbanding issue manifests in spectral windows created using the frequency division multiplexing (FDM) mode of the ALMA Baseline Correlator (BLC) \citep{2007A&A...462..801E, cortes_2022_7764458}. In Cycle 9 there was no heuristic in the ALMA pipeline to detect subbanding. As of Cycle 12 a quantitative heuristic has been introduced, discussed in Section~\ref{sec:heuristic}.

\begin{figure}[t!]
\begin{center}
\includegraphics[width=0.5\textwidth]{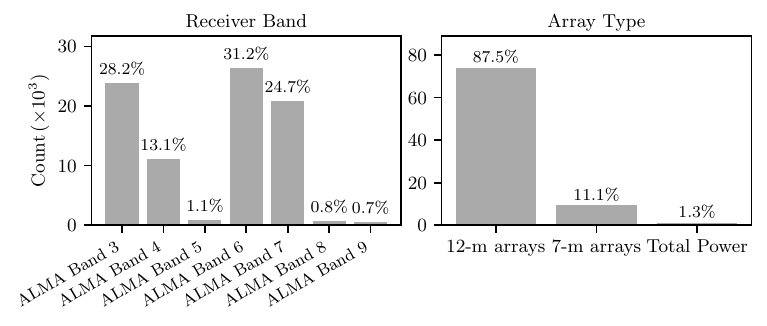}
\caption{Histograms of categorical features by bandpass solutions in the selected Cycle 9 training dataset.}
\label{fig:C9-dataset}
\end{center}
\end{figure}

The Cycle 9 data set contains 84,572 bandpass calibration solutions from 303 execution blocks (EBs), belonging to 151 member observing unit sets (MOUSs). Each MOUS contains one or more EB(s), and all data within a MOUS are merged into a single set of science products, e.g. image or spectral cube, \citep{cortes_2022_7764458}. The distribution of categorical features of this Cycle 9 data set is shown in Figure~\ref{fig:C9-dataset}, covering six ALMA receiver bands, 12-m array and 7-m array observations. 

A second data set, used to evaluate the generalizability of the classifier framework, was assembled from the recent public ALMA Cycle 11 observations with project codes starting with $\#$2024 (October 2024 -- September 2025). Beginning in late Cycle 10, ALMA began automated data delivery (based on QA scores), and Cycle 11 is the most recent cycle for which nearly all data are publicly available. We restricted to Cycle 11 calibration data that were inspected by data reducers during the ALMA QA process. It contains 744,240 bandpass calibration solutions from 2,641 EBs, belonging to 1,085 MOUSs, covering ten ALMA receiver bands (Band 1, and Band 3--10). Different from the main Cycle 9 dataset, the 7-m array observations in Cycle 11 were processed through the BLC \citep{cortes_2025_14933753}, whereas previous cycles used the Atacama Compact Array (ACA) correlator.

\subsection{Additional Metadata}

At ALMA observing frequencies, strong atmospheric spectral lines, mainly from the water vapor and ozone, may introduce absorption features in the bandpass solutions at specific transition frequencies. While atmospheric transmission effects can be corrected using the Atmospheric Transmission at Microwaves (ATM) code implemented in the ALMA pipeline \citep{2001ITAP...49.1683P}, these features in the bandpass amplitude arrays can spuriously trigger the anomaly detector and bias the classification; such features manifest as drops in amplitude within the bandpass spectrum, which can appear similar to the platforming effects to be caught, as shown in Figure~\ref{fig:anomalies_example}. To characterize these atmospheric features, we constructed an atmospheric transmission model at 1 MHz resolution based on the ATM code. For each of the ALMA's ten frequency bands, we constrained the transmission value curve by gas pressure, altitude, and precipitable water vapor (PWV), quantities which for our purposes are well-known and regularly measured at the ALMA Array Operations Center. The PWV values for each band were chosen based on the conditions in which that band typically observes, and the resultant set of curves were concatenated into a single representative all-band model. We then ran a standard peak-finding algorithm on the full curve to identify and label atmospheric line locations and their widths, in frequency, which defines our regions of atmospheric interference. These regions were then passed to the anomaly detector in order to help reduce false positives due to atmospheric interference.

\section{Methodology \label{sec:meth}}
We developed a supervised classifier framework that learns the mapping from features extracted from bandpass calibration solutions to Boolean anomaly labels. This framework is general with respect to feature construction and label definition, which can be based on either the amplitude or phase portion of complex bandpass solutions, or both. The current implementation considers only anomalies in the amplitude solutions. As the raw input data, bandpass amplitude gains are 1D arrays of length varying across EBs, and are typically normalized to a mean of 1.0. Because most standard ML classifiers require fixed-length input vectors, the raw data arrays are flattened into a set of numerical representations, i.e., feature vectors. As the predictive model, we use eXtreme Gradient Boosting \citep[XGBoost, ][]{DBLP:journals/corr/ChenG16}, which builds an additive ensemble of decision trees sequentially via boosting.

\subsection{Labels \label{sec:lables}}

\begin{figure*}[th!]
    \begin{center}
    \includegraphics[width=\textwidth]{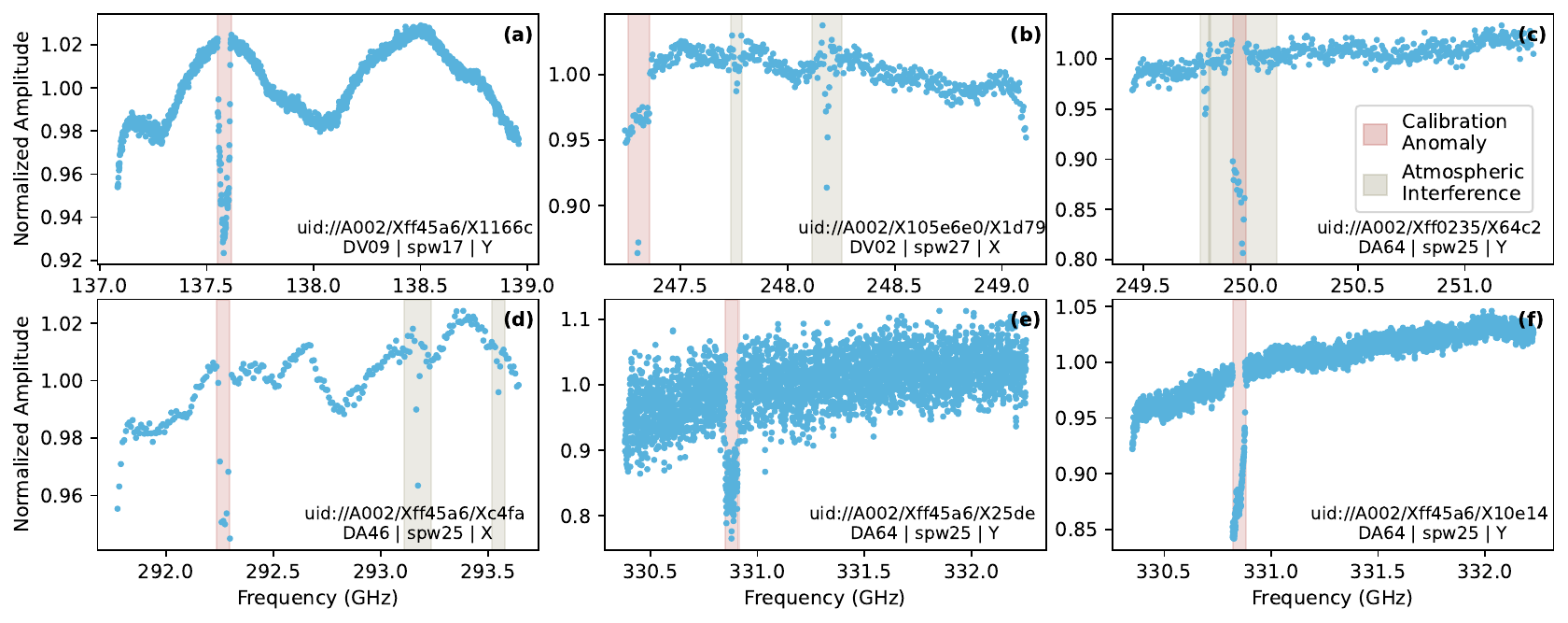}
    \caption{Example individual bandpass amplitude solutions, with anomalous intervals shaded in red and non-problematic atmospheric interferences shaded in brown. Panel~(c) shows an instance where a platforming anomaly intersects with the atmospheric interference (Section~\ref{sec:data}).}
    \label{fig:anomalies_example}
    \end{center}
\end{figure*}

In this work, we define a bandpass calibration anomaly as an interval in the bandpass amplitude solution that deviates from the local background shape, where the associated bandpass solution requires the flagging action. Figure~\ref{fig:anomalies_example} shows example amplitude solutions with anomalous intervals. This definition matches the practical calibration procedure and captures a broad class of anomaly types, among which the subband artifact is a subset that accounts for a large fraction of the observed cases.

We adopt a Boolean labeling scheme, with \texttt{True} representing solutions having bandpass amplitude anomalies and \texttt{False} representing solutions without such anomalies. The ground-truth labels used for model training were defined from a combination of the ALMA Quality Assurance level-2 (QA2) flagging action and additional label review by the authors with ALMA calibration expertise. During the ALMA QA2 process, data reducers inspect the data and flag anomalous data with documented QA2 flagging reasons, with each reason corresponding to a specific task in the ALMA Science Pipeline \citep{2023PASP..135g4501H}. A detailed description of the ALMA QA process is provided in \citet{cortes_2022_7764458}. $\sim4.4\%$ of the selected Cycle 9 dataset (at the solution level) was manually flagged during the QA2 review, and $\sim0.6\%$ has manual flagging reasons related to bandpass amplitude anomalies\footnote{\footnotesize{The relevant QA2 flagging reasons include ``\texttt{bandpass\_amplitude\_frequency}" and ``\texttt{bandpassflag\_amplitude\_frequency}"; flags related to system temperature behavior were excluded.}}. These QA2 flagging reasons were used to effectively define the initial \texttt{True} labels. We further validated and, where needed, revised the labels.

Looking forward, labels derived directly from the characteristics of the calibrated visibilities may better represent the issues that adversely impact science products. Exploring more scalable ways to identify these issues, with less reliance on human generated labels, will be the subject of future work.

\subsection{Features \label{sec:feat}}

\begin{deluxetable*}{cll}
    \tablecaption{Constructed Features \label{tab:features}}
    \tabletypesize{\scriptsize}
    \tablewidth{0pt}
    \setlength{\tabcolsep}{4pt}
    \renewcommand{\arraystretch}{1.15}
    \tablehead{
    \colhead{Features} & \colhead{Description} & \colhead{Feature construction}}
    \startdata
        Masked score ($S_{\rm masked}$) & Anomaly score for a spectral  & Derived using the scan statistic with both $5\%$ spectral \\
        \nodata & interval deviating from the baseline & edges and atmospheric interferences masked;\\
        \nodata & & dimensionless\\
        Anomaly width for masked score ($W_{\rm masked}$) & Frequency range of the spectral & Same as masked score; units of Hz  \\
         &interval with the highest masked score  & \nodata\\
        Unmasked score ($S_{\rm unmasked}$) & Anomaly score for a spectral & Derived using the scan statistic with $5\%$ spectral \\
        \nodata & interval deviating from the baseline & edges masked, but atmospheric interferences unmasked;\\
        \nodata & \nodata & dimensionless\\
        Anomaly width for unmasked score ($W_{\rm unmasked}$) & Frequency range of the spectral interval & Same as unmasked score; units of Hz  \\
         & with the highest unmasked score & \nodata\\
        Fixed-width score ($S_{\rm fixed-width}$) & Anomaly score for a spectral & Derived using the scan statistic with both $5\%$ spectral \\
        \nodata & interval deviating from the baseline & edges and atmospheric interferences unmasked;\\
        \nodata & \nodata & fixed interval width of 62.5 MHz;\\
        \nodata & \nodata & dimensionless\\
        $\texttt{NormalizedDispersion}^{(4)}$ & Bandpass amplitude dispersion & Computed using Equation~\ref{eqa:dispersion} from amplitude array\\
        \nodata & \nodata & with a channel separation of 4; dimensionless\\
        Channel spacing ($\Delta f$)& Effective spectral interval  & Computed from frequency array;\\
        \nodata & \nodata & units of Hz\\
        ALMA receiver band & Observational frequency band & Categorical feature \\
    \enddata
    \tablecomments{The selected feature set, used for the XGBoost model, includes seven statistical features flattened from the bandpass amplitude solution and a categorical feature, see Section~\ref{sec:feat}.}
\end{deluxetable*}

XGBoost excels at building predictors from combinations of different types of features. We constructed two types of features: statistical and categorical. The statistical ones include metrics based on interval anomaly detection methods applied to several levels of cleaning the data, normalized array dispersion, and channel spacing. For the categorical features, we need to pay particular attention to potential data leakage -- not because it would make our calculated performance worse, but because it would make the evaluation less reliable.  In particular, we do not want to encode dataset-specific associates with the labels. For example, bandpass anomalies in the training data were associated with specific antennas during ALMA Cycle 9 observations. This association was cycle-specific because it resulted from the temporary instrumental conditions of that cycle. Including antenna name as features could therefore allow the classifier to memorize this dataset-specific pattern between antenna and anomaly labels, resulting in reduced generalization.

The implemented features are discussed below, with the full set of features summarized in Table~\ref{tab:features}. Beyond the presented implementation, the framework can be extended to include additional features derived from the complex bandpass solutions, not only amplitude-frequency behavior but also phase-frequency behavior, with corresponding updates to the labeling strategies.

\subsubsection{Scoring metrics based on NWKR scan statistics}
The primary features in our classifier are based on an interval anomaly detection method called ``Scan Statistics'' \citep{Kulldorff1997, 2024hss..book.....G}. The classic scan statistics method assumes clean and well-behaved data other than some potentially anomalous region. \citet{rakib2026efficientregressionmodelsscan} adapts this to the more complicated characteristics observed in ALMA Bandpass solutions by modeling the background with a Nadaraya-Watson kernel regression (NWKR) model, and then scans all possible intervals looking for the one that most deviates from that fit. It identifies a region which is quantiatively ``most anomalous'' and provides a score. The score ranges from $0$ to $1$, with large values representing a more anomalous interval. This score is provided for each interval, and the score of the largest scoring interval called a \emph{scan statistic} is returned, along with the corresponding interval.  

As this scan statistic method is designed for data that (without an interval anomaly) is well-fit by NWKR, it is not always appropriate for the real calibration data. As a result, we run it in three different ways, producing different features with each fed into XGBoost.  

The first features are a score $S_{\rm masked}$ and found interval width $W_{\rm masked}$, which we refer to as \emph{masked}. These only consider intervals that do not overlap with known atmospheric lines; that is, it masks out the lines with known atmospheric interference. Otherwise, for a given amplitude solution that has atmospheric lines, the scan statistic tends to identify intervals in the atmospheric interference range, biasing the scoring metric and resulting in false anomalies. Therefore, anomaly scores computed on all channels are typically higher than when the channels affected by the known atmospheric lines are excluded. Masking the affected channels is effective in isolating atmospheric lines from instrumental calibration anomalies and reducing false detections.

The second set of features are the same score $S_{\rm unmasked}$ and found interval width $W_{\rm unmasked}$, calculated without the masking of atmospheric lines. This is sometimes necessary for when platforming occurs in intervals that overlap with atmospheric interference, such as in Figure~\ref{fig:anomalies_example} panel (c). In addition to channels flagged by other calibration steps, both $S_{\rm masked}$ and $S_{\rm unmasked}$ exclude the first and last $5\%$ of channels in each spectral window.

The last scan-statistic feature is the score $S_{\rm fixed-width}$, calculated using a restricted form of the scan statistic method -- one in particular that fixes the interval width at 62.5 MHz. This corresponds to the width of the BLC FDM subband, and this width is characteristic of the platforming anomalies observed in the ALMA BLC. This size can be adapted to reflect different hardware setups, supporting generalizability across facilities (e.g. extending to 200 MHz for the ALMA WSU). \(S_{\rm fixed-width}\) excludes only channels flagged by other calibration steps and otherwise considers all spectral channels without masking atmospheric lines.

\subsubsection{Additional Expert-Informed Features}

We introduced several, additional expert-selected features to represent other salient characteristics of the data. As the only categorical feature, we include the ALMA receiver band, with each band mapping to a fixed frequency range (Figure~\ref{fig:C9-dataset}). This feature provides context for frequency-dependent effects, including atmospheric transmission, rather than correlator behavior. We also include the channel spacing, in Hz, of amplitude solutions. It represents the effective spectral resolution, determined by the inherent correlator resolution and bandwidth setup, together with the smoothing function and channel-averaging factor applied during calibration.

In addition, we included another statistical feature to measure the amplitude dispersion, we used the normalized median absolute deviation (MAD, ${\rm median}(|X_i -{\rm median}(X)|)$) of the fourth-order difference, 
\begin{equation}
{\rm NormalizedDispersion}^{(4)} = \frac{{\rm MAD}(A_i - A_{i+4})}{{\rm MAD}(A_i)} \label{eqa:dispersion},
\end{equation}
where $A_i$ is the bandpass calibrator amplitude at channel $i$. We use a channel separation of $4$ because of the non-negligible correlation observed between neighboring channels (e.g. $i$ and $i+1$) introduced by the smoothing function of ALMA’s correlators (i.e. Hanning smoothing). This channel correlation, however, becomes effectively negligible at a separation of four channels, allowing the feature to be a reliable estimate of independent amplitude variations \citep{2018AJ....155..182L}. Normalization by the overall dispersion of each solution, ${\rm MAD}(A_i)$, improves comparability across solutions.

\subsection{Data Preparation \label{sec:data_prep}}

In order to avoid introducing spurious polarization-dependent effects in the data, both polarizations of a given antenna and SPW are always flagged together, even when the issue only occurs in one polarization. While necessary from a data-reduction perspective, this procedure introduces inconsistencies in solution-level labeling. To account for this, we aggregate solutions at the polarization-pair level and combine their labels using a logical \texttt{OR} operator.

For feature construction at the polarization-pair level, the six statistical features are derived independently from the amplitude array of each polarization, while channel spacing and the categorical ALMA receiver band are common to both polarizations. As such, each polarization-pair aggregation has 14 features in total.

\subsection{Training Strategy \label{sec:training_setup}}

Because most ALMA data are of high quality, the training dataset is highly imbalanced. Following polarization-pair aggregation, bandpass amplitude anomalies account for only $\sim0.5\%$ of the 42,286 solution pairs. This imbalance reflects the intrinsic distribution of real observational data, where anomalous calibration solutions are rare compared to nominal solutions. To account for the imbalanced class, we use stratified sampling for the train-test split and give more weight to the minority \texttt{True} class during model training.

Train and test splits were performed at the EB level to avoid information leakage between training and evaluation sets. For each train/test split, a given EB (and thus all its bandpass solutions) is assigned to either the training, validation, or test set. Because solutions from the same EB typically share similar observing and instrumental conditions, this prevents non-independent solutions from existing in both the training and evaluation sets and thus prevents overestimation of model performance. Moreover, although EB IDs are not included as model inputs, they may be abstractly encoded in other input features. Splitting at the EB level ensures that, even if the model learns EB-specific patterns during training, the evaluation on the test set with previously unseen EBs remains trustworthy.

\section{Evaluating Effectiveness \label{sec:eva}}

\subsection{Nested Cross-Validation}

\begin{figure*}
\begin{center}
    \includegraphics[width=0.9\textwidth]{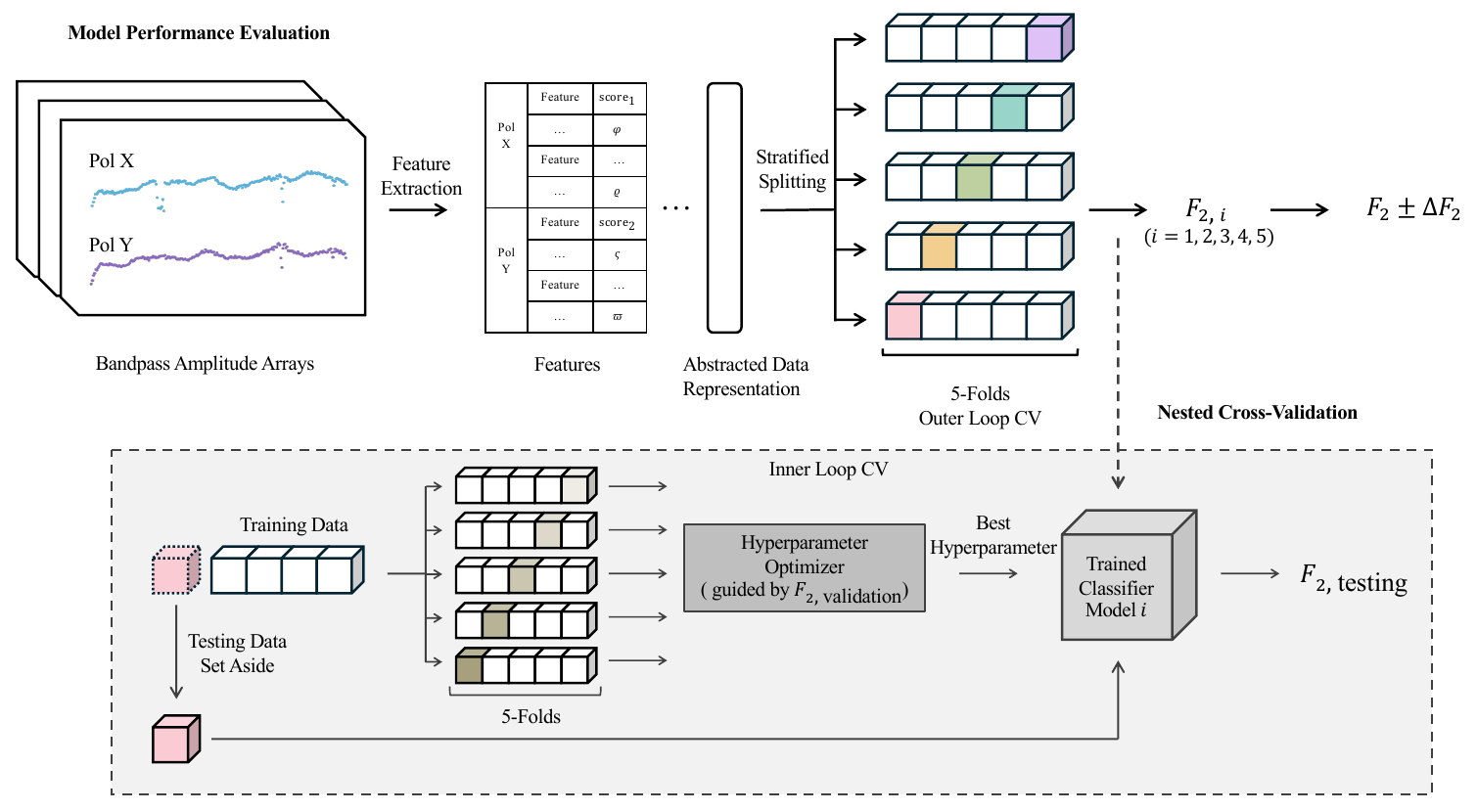}
    \caption{Schematic of the classifier evaluation framework workflow using nested cross-validation. The 5-fold cross-validation procedure produces five optimized hyperparameter sets and corresponding models, which together estimate the range of $\mathrm{F_2}$ values and classifier performance. Colored blocks (purple, lime, green, orange, and pink) represent the held-out outer test folds, brown blocks represent the held-out inner validation folds, and white blocks represent the training folds used in each iteration.}
    \label{fig:cv_schematic}
\end{center}
\end{figure*}

The XGBoost classifier hyperparameters include maximum depth of a tree, learning rate, number of estimators, and subsample ratio of the training instances. We optimized these hyperparameters using nested cross-validation with the $\mathrm{F}_2$ score as the optimization metric:
\begin{equation}
\mathrm{F_2}
= \frac{\beta^2+1}{(\beta^2\cdot\rm{recall}^{-1}) + \rm{precision}^{-1}},
\end{equation}
where $\beta=2$, $\mathrm{recall}=\mathrm{TP}/(\mathrm{TP}+\mathrm{FN})$ is the fraction of true anomalies identified, and $\mathrm{precision}=\mathrm{TP}/(\mathrm{TP}+\mathrm{FP})$ is the fraction of predicted anomalies that are true anomalies\footnote{\footnotesize TN, TP, FP, and FN denote true negatives (calibration solution not labeled and not identified by classifier), true positives (calibration solution labeled as anomalous and identified by classifier), false positives (calibration solutions not labeled as anomalous but identified by the classifier), and false negatives (calibration solutions labeled as problematic but not identified by the classifier), respectively.}. We use a value $\beta=2$ so that the $\mathrm{F}_2$ score weighs false negatives more, aligning with the objective of anomaly detection to prioritize identifying true anomalies over minimizing additional false positives. In addition, we found that the measured $\mathrm{F}_2$ score and model performance are sensitive to the train-test split. This is because anomaly examples are unevenly distributed across EBs, and therefore different train-test splits produce test sets with anomalous cases of varying difficulty. As such, in contrast to a single train-test split, we used nested 5$\times$5 cross-validation, as illustrated in Figure~\ref{fig:cv_schematic}. In each outer fold, $80\%$ of the EBs were used for training and validation and $20\%$ were held out for testing. Each of the five outer folds serves as the test set once, allowing every labeled anomaly to be evaluated exactly once. Within each outer-training set, five-fold inner cross-validation is used for hyperparameter optimization. The outer folds assess model performance and its variation across splits.

\subsection{Results}

\begin{deluxetable*}{clllll}
    \tablecaption{Performance Metrics Across Outer Test Folds for Different Feature Sets\label{tab:outerfolds}}
    \tabletypesize{\scriptsize}
    \tablewidth{0pt}
    \tablehead{
    \colhead{Feature Set} & \colhead{TN} & \colhead{TP} & \colhead{FP} & \colhead{FN} & \colhead{$\overline{F_2} \pm \Delta F_2$}\\
    & \multicolumn{4}{c}{Total Count (Average Percentages)\tablenotemark{$^\ast$}} & }
    \startdata
        [$S_{\rm masked}$] & 41772 ($98.79\% \pm 0.18\%$) & 163 ($0.39\% \pm 0.06\%$) & 310 ($0.73\% \pm 0.21\%$) & 41 ($0.10\% \pm 0.04\%$) & $0.63 \pm 0.10$ \\{}
        [$S_{\rm masked}$, $W_{\rm masked}$] & 41980 ($99.28\% \pm 0.09\%$) & 157 ($0.37\% \pm 0.06\%$) & 102 ($0.24\% \pm 0.12\%$) & 47 ($0.11\% \pm 0.05\%$) & $0.73 \pm 0.10$ \\{}
        [$S_{\rm masked}$, $W_{\rm masked}$, $S_{\rm unmasked}$, $W_{\rm unmasked}$, $S_{\rm fixed-width}$] & 42039 ($99.42\%\pm 0.06\%$) & 176 ($0.42\% \pm 0.04\%$) & 43 ($0.1\% \pm 0.08\%$) & 28 ($0.07\% \pm 0.04\%$) & $0.85 \pm 0.05$ \\{}
        [$S_{\rm masked}$, $W_{\rm masked}$, $S_{\rm unmasked}$, $W_{\rm unmasked}$, $S_{\rm fixed-width}$ & \multirow{2}{*}{42055 ($99.45\% \pm 0.04\%$)} & \multirow{2}{*}{180 ($0.43\% \pm 0.03\%$)} & \multirow{2}{*}{27 ($0.06\% \pm 0.05\%$)} & \multirow{2}{*}{24 ($0.06\% \pm 0.02\%$)} & \multirow{2}{*}{$ 0.88\pm0.02$} \\{}
        ${\texttt{NormalizedDispersion}^{(4)}}$, $\Delta f$, ALMA receiver band] & & & & &
    \enddata
    \tablecomments{Evaluation was performed on the ALMA Cycle 9 training dataset, which contains 84,572 individual bandpass calibration solutions corresponding to 42,286 polarization-pair samples. $\ast$: Confusion matrix counts summed over the five outer test folds and the average normalized confusion matrix (per-fold percentages).}
\end{deluxetable*}

The combination of XGBoost together with the scan statistics performed conjunction operations to understand the data distribution. As the baseline feature set, using only $S_{\rm masked}$ from the scan statistic \citep{rakib2026efficientregressionmodelsscan} obtains an $F_2$ score of $0.63 \pm 0.10$ across the five outer test folds. We extend this baseline by including the corresponding interval width, $W_{\rm masked}$, followed by the unmasked and fixed-width features to catch more scenarios of anomalies, and pinpoint the precise nature of the platforming issue. As shown in Table~\ref{tab:outerfolds}, iteratively adding these features improves the classifier performance. Allowing XGBoost to also train on these obtains an increasing mean $F_2$ to $0.85 \pm 0.05$ with reduced fold-to-fold deviation, measured across the same folds.

Because of the high imbalance between the \texttt{True} and \texttt{False} classes (i.e. the anomalies and nominal data), XGBoost's ability to learn the less common anomaly class is limited. Therefore, we further included additional expert-informed features that explicitly address radio-specific complications. With the full set of eight features (six features computed separately for each polarization and two features common to both), the XGBoost classifier leveraged complementary domain knowledge to improve performance beyond the scan statistic features. The addition of these expert-informed features further improves the mean $F_2$ score to $0.88 \pm 0.02$ across the outer test folds.

\begin{figure}
\begin{center}
    \includegraphics[width=\columnwidth]{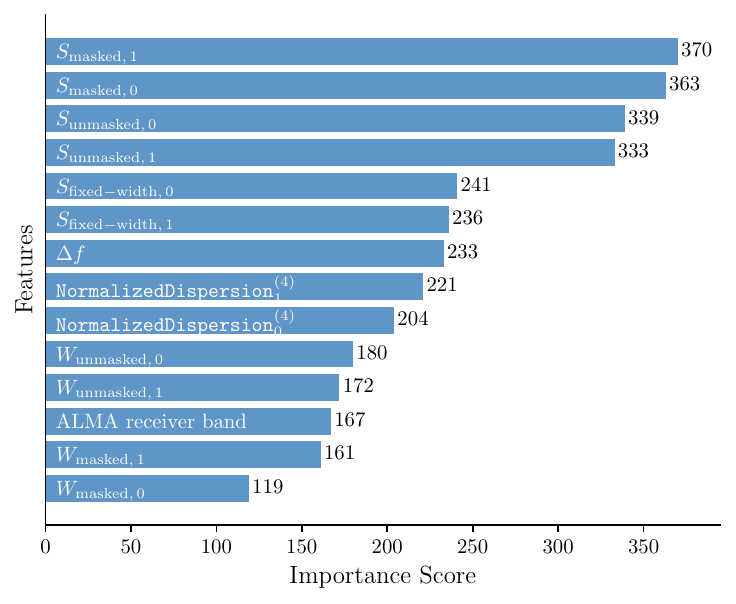}
    \caption{Feature importance of the implemented features by weight, measured by the number of times a feature is used for a split across all trees in the model. Subscripts (0, 1) correspond to features derived separately from the two orthogonal polarization solutions; features common to both polarizations have no subscript (see Section~\ref{sec:data_prep}).}
    \label{fig:importance}
\end{center}
\end{figure}

In the full-feature model, $S_{\rm masked}$ shows a strong influence and is closely related to the model’s decision process (Figure~\ref{fig:importance}). Because there may be additional features not included in the used feature set that are correlated with features having high importance scores, a detailed analysis of the causes of the anomaly predictions is beyond the scope of this work.

\subsection{Comparison to ALMA pipeline heuristic \label{sec:heuristic}}

\begin{figure*}
    \centering
    \gridline{
        \fig{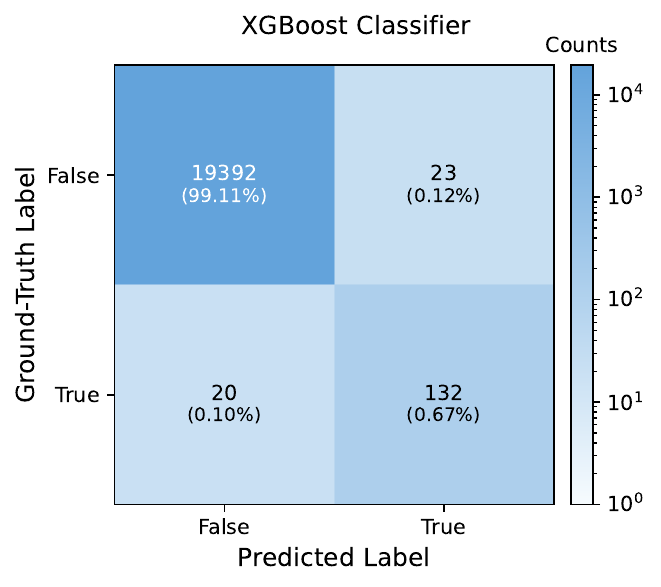}{0.5\textwidth}{}
        \fig{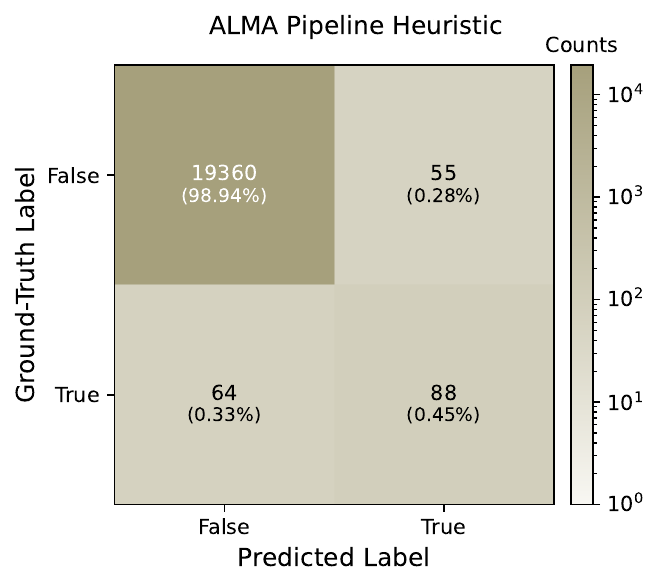}{0.5\textwidth}{}
    }
    \caption{\label{fig:confusion_matrix_C9} 
        Confusion matrices for the machine learning classifier (blue color scale, left) and the ALMA calibration pipeline heuristics (brown color scale, right), respectively. Model predictions were compared against ground-truth labels derived from the combination of ALMA QA2 flagging and expert review. Because the pipeline heuristic applies only to the observations in the FDM operating mode of the BLC, the comparison is restricted to the 19,567 polarization-pair samples in the Cycle 9 training dataset for which predictions from both methods are available.}
\end{figure*}

The bandpass subband heuristic implemented in the current ALMA pipeline evaluates anomalous features associated with a subband width of 62.5 MHz, using five threshold-based heuristics \citep{almapipeline2025}. It is designed to identify specifically correlator subband issues (including platforming), and applies only to spectral windows processed in the FDM operating mode of the BLC. Accordingly, the selected Cycle~9 training dataset contains 39,134 applicable FDM bandpass solutions. Following the same evaluation framework (Section~\ref{sec:data_prep}), we aggregated the prediction from the two orthogonal polarization bandpass solutions, and obtained 19,567 polarization-pair samples. Compared against the ground-truth labels defined in Section~\ref{sec:lables}, the confusion matrices of the XGBoost classifier and the current pipeline heuristic are shown in Figure~\ref{fig:confusion_matrix_C9} (left and right, respectively). The classifier substantially improves the performance for detecting bandpass amplitude anomalies, reducing both the false-negative rate (FNR) and false-positive rate (FPR)~\footnote{\footnotesize We use the standard statistical definitions, \(\mathrm{FNR}=\mathrm{FN}/(\mathrm{TP}+\mathrm{FN})\) and \(\mathrm{FPR}=\mathrm{FP}/(\mathrm{TN}+\mathrm{FP})\), different from the ALMA internal convention of normalizing by all records.}. Over the 19,567 polarization-pair samples, the FNR decreases from $42\%$ for the heuristic to $13\%$ for the classifier, while the FPR decreases from $0.3\%$ to $0.12\%$. Although the FPRs of both methods are low, ALMA processing aggregates the results from many independent assessments in determining which datasets need manual review, so a modest FPR for individual assessments can lead to a substantial human inspection workload.

We consider that two elements account for the performance improvement of the ML classifier, relative to the current ALMA pipeline heuristic. First, the XGBoost classifier learns and optimizes complex, non-linear decision boundaries across multiple features. In contrast, the pipeline’s current QA scoring framework treats quantitative features independently, each with a univariate decision boundary, and classifies a sample as anomalous if any single feature crosses the threshold. Second, the implemented scan-statistic feature set \citep {rakib2026efficientregressionmodelsscan} is particularly well-matched to the target subband anomalies.

\section{Discussion \label{sec:disc}}

\subsection{Limitation \label{sec:limit}}

We examined misclassified calibration solutions in the test folds, including false negatives (Figure~\ref{fig:fn}), corresponding to anomalous solutions missed by the classifier; and false positives (Figure~\ref{fig:fp}; Appendix~\ref{apx:misclassification}), corresponding to false alarms. One class of false negatives involves the platforming anomalies at the edge of amplitude solutions (Figure~\ref{fig:fn}~(1)--(4) and (6)--(7)). 

Because the digital filter response leads to a loss of sensitivity at both edges of spectral windows, we constructed the central scan statistics (masked and unmasked) to exclude the extreme edge channels of the amplitude solutions (the first and last $5\%$ of an array). As such, a substantial region of an edge platforming anomaly was not evaluated by the scan statistic, limiting the classifier's ability to identify these anomalies. We explored additional features designed to identify edge platforming, including searches within $10\%$ of each end of the spectrum for step-like discontinuities and their overlap with the fixed-width scan statistic intervals. However, these features did not improve classifier performance so were not included in the final feature set. Moreover, one false negative involves two anomalous intervals, located at the edge and in the middle respectively (Figure~\ref{fig:fn}~(1)), whereas the scan statistic assumes a single anomalous interval per solution. 

Further review found that the false negatives in Figure~\ref{fig:fn}~(11)--(22) are related to an issue caused by an incorrectly seated correlator card. This issue produced significant subband platforming in several antennas within the same EBs, which were correctly identified by the classifier, but also introduced artifacts in the solutions for other antennas. In particular, the false negatives in Figure~\ref{fig:fn}~(16)--(22) are associated with a specific EB, which was affected by subbanding in the reference antenna. Since antenna-based calibrations are relative to the reference antenna, several other antennas were corrupted by the anomaly in the reference antenna, showing unusual morphologies, and were consequentially missed by the classifier. These corrupted antennas were flagged by data reducers so show up as false negatives, although they would likely not have been anomalous had the reference antenna been good. The classifier identified the directly affected antennas, suggesting that these false negatives in Figure~\ref{fig:fn}~(11)--(22) are propagated calibration artifacts rather than missed subband anomalies. A definitive determination of whether these propagated artifacts are consequential anomalies can be made upon assessing the calibrated data products.

The false positives identified by the classifier are shown in Figure~\ref{fig:fp}, with the anomalous intervals shaded in red. Some false positive solutions happen to contain spikes introduced by the WVR Local Oscillator (LO) leakage, which occurs at frequencies corresponding to the (sub)harmonics of the 183.3~GHz LO signal, such as the spikes at 91.66 GHz in Figure~\ref{fig:fp}~(22)--(24). However, these solutions were not classified as anomalous because of the LO leakage artifacts. Although the classifier, unlike the current pipeline heuristics, does not explicitly exclude the known WVR LO frequency ranges, the anomalous intervals identified by the scan statistic do not coincide with such frequencies. Moreover, the incidence of these leakage artifacts is low in the current training dataset. We therefore find no clear evidence that implementing a veto for the LO leakage frequencies would change these misclassifications. Dedicated mitigation should be evaluated in future work.

We have assumed a relatively simple ground-truth rule as a prediction target for our classifier: a specific class of bandpass calibration anomalies should have the affected antenna and spectral window flagged. A more precise determination of the circumstances under which these anomalies, or the classifier misclassifications, if not flagged or otherwise mitigated, adversely impact the quality of the downstream calibrated data will be the subject of future work.

\subsection{Application to Cycle 11 Observations \label{sec:c11}}

To assess generalizability to unseen observations, we applied the classifier model trained on the Cycle 9 dataset to a per-cycle testbed of Cycle 11 observations. This Cycle 11 dataset is $\sim\,9$ times larger than the Cycle 9 training dataset, containing 372,120 solution pairs (744,240 individual solutions). The Cycle 11 dataset had two notable characteristics that were not in the Cycle 9 training data: Cycle 11 had a new receiver band, Band 1, that was not offered for PI science during Cycle 9; and the 7-m array data were correlated with the BLC, not the ACA correlator used in Cycle 9 and earlier cycles. Note that the target bandpass anomalies are only present in BLC data, so will not have been represented in the Cycle 9 7-m training data.

It has not yet been feasible to carefully review the QA2 flagging by data reducers for this dataset to ensure consistent labeling, as was done for the Cycle 9 training dataset. For this reason, we have not used QA2 flagging as a source of ground truth, but instead compare the classifier predictions with the QA2 flagging actions. For the 12-m array observations, the classifier predicted 566 solution pairs to have bandpass amplitude anomalies, while data reducers manually flagged 849 pairs for QA2 reasons related to bandpass amplitude anomalies. Of these, 209 pairs were identified by both the classifier and data reducers. We subsequently reviewed the 1206 unique pairs identified by either method and confirmed 335 to have meaningful subband anomalies: 205 of the 209 overlapping (classifier/data reducer flags) pairs and 130 identified by only one of the two methods. For Band 1 observations, the classifier identified 28 pairs, 27 of which were confirmed to have meaningful anomalies; 26 of these 27 were also caught during the routine QA2 process. These Band 1 detections suggest generalization to a new observing band.

About three quarters of the bandpass solutions flagged by data reducers but identified as \texttt{False} by the classifier are related to WVR LO leakage artifacts in Band 1. The flagging guidelines for LO leakage artifacts have varied across ALMA observing cycles. The recommended procedure in Cycle 11 differed from that in Cycle 9: a WVR LO harmonic strongly affects Band 1 at 45.83 GHz, close to the methanol maser transition at 45.84 GHz, leading to a recommendation to flag LO leakage artifacts at this frequency if a science line is also detected. Given the varying labeling policy for such artifacts --- as well as the dependence on science data not present here --- it is not unexpected that they are not identified by the classifier. Since the leakage artifacts occur at known frequencies, we regard them as low priority anomalies for the ML classifier. It could be considered in the future, with a consistently labeled dataset containing LO spikes, to optimize the classifier's ability to handle them.

For the 7-m array observations, the classifier predicted 109 solution pairs to be positive. However, we did not find clear subband anomalies in the BLC-processed 7-m data. Most of these positive predictions occur in amplitude solutions with very few channels ($<20$) and high noise. With so few channels, identifying an anomaly from random (and high) variation becomes difficult unless the anomaly is particularly obvious. Similar results are also present in the Cycle 9 FP examples (Figure~\ref{fig:fp}~(6)--(7)). Further expansion of the training dataset might improve classifier performance for solutions with high noise and limited channel numbers.

Because we reviewed only solutions identified positive by either the classifier or data reducers or both, we cannot assess anomalies among solutions classified as negative by both methods. We therefore estimated lower bounds on the classifier FNR and FPR. Across the 372,120 Cycle 11 polarization-pair samples, the classifier predicted 299 TP, 376 FP, and $\geq36$\,FN, relative to our review. This confusion matrix corresponds to lower bounds of FNR\,$\geq 11\%$ and FPR\,$\geq 0.1\%$, consistent with the performance across the five outer test folds of the Cycle 9 training dataset.

\section{Summary \label{sec:sum}}
We developed a supervised classifier framework for distinguishing anomalies from benign variation in ALMA bandpass calibration solutions. The framework is general with respect to feature construction and anomaly types. It can be extended to include additional features derived from the amplitude and/or phase portions of complex bandpass solutions, and to target amplitude and/or phase anomalies.

The current implementation focuses on amplitude anomalies. The classifier was trained on selected ALMA Cycle 9 observations containing 84,572 bandpass calibration solutions. Ground-truth labels were defined from a combination of ALMA QA2 flagging actions and additional label review. We used eight features derived from NWKR scan statistics and additional expert-informed features that account for radio-specific complications, including atmospheric interference, characteristic subband width, and amplitude dispersion. These features were used as inputs to the XGBoost classifier, which builds an additive ensemble of decision trees to produce predictions. Individual solutions were paired by polarization, with polarization pairs treated as the unit of classification.

We evaluated the classifier using nested $5\times5$ cross-validation with EB-level splitting to avoid information leakage between training and evaluation data. Performance was assessed from the five outer test folds using $F_2$ and confusion-matrix statistics. Iteratively adding statistical anomaly metrics and domain knowledge improved classification beyond true baseline scan-statistic score alone. This helps the classifier capture more scenarios of anomalies, and pinpoint the nature of the subband issue. With the full feature set, we achieved a mean $F_2$ score of $0.88\pm0.02$ across the five outer test folds. Compared with the current ALMA pipeline subband heuristic on the applicable FDM subset, the classifier reduced the false-negative rate from $42\%$ to $13\%$ and the false-positive rate from $0.3\%$ to $0.12\%$.

We further applied the trained classifier to 744,240 bandpass calibration solutions from ALMA Cycle 11 observations, nearly an order of magnitude larger than the training dataset. These Cycle 11 data include configurations not represented in the Cycle 9 training dataset, including Band 1 observations. The classifier identified meaningful Band 1 anomalies that should have been flagged, suggesting that the framework can generalize across unseen observations and different observing bands.

The remaining misclassifications note limitations of the current feature representation, including edge platforming, multiple anomalous intervals within a single solution, and solutions with high noise levels and limited channel numbers. Future work will assess whether misclassified calibration solutions propagate into calibrated data products and affect science data quality.

\section{Data access \& code}
The training dataset of the selected ALMA Cycle 9 bandpass calibration solutions is publicly available in the Zenodo Repository \citep{xue_2026_21972118}. The code \texttt{Bandpass Amplitude Anomaly Classifier} is open source and available at \url{https://github.com/cixue/bandpass-classifier} with the model trained on the Cycle 9 data using the full feature set.


\facilities{ALMA}

\software{
    \texttt{XGBoost} \citep{xgboost},
    \texttt{Scikit-learn} \citep{scikit-learn},
    \texttt{bandpass-classifier},
          }

\acknowledgments
This work was supported by the National Science Foundation under Cooperative Agreement 2421782 and the Simons Foundation grant MPS-AI-00010515 awarded to the NSF-Simons AI Institute for Cosmic Origins - CosmicAI, https://www.cosmicai.org/. This paper makes use of public ALMA calibration data and observation metadata, but not the scientific data, from all projects starting with: ADS/JAO.ALMA\#2024. The National Radio Astronomy Observatory is a facility of the National Science Foundation operated under cooperative agreement by Associated Universities, Inc. ALMA is a partnership of ESO (representing its member states), NSF (USA) and NINS (Japan), together with NRC (Canada), MOST and ASIAA (Taiwan), and KASI (Republic of Korea), in cooperation with the Republic of Chile. The Joint ALMA Observatory is operated by ESO, AUI/NRAO and NAOJ.

\bibliography{xue}

@manual{cortes_2022_7764458,
  title        = {ALMA Cycle 9 Technical Handbook},
  author       = {Cortes, Paulo and
                  Remijan, Anthony and
                  Hales, Antonio and
                  Carpenter, John and
                  Dent, Bill and
                  Kameno, Seiji and
                  Loomis, Ryan and
                  Vila Vilaro, Baltasar and
                  Biggs, Andy and
                  Miotello, Anna and
                  Vlahakis, Catherine and
                  Rosen, Rachel and
                  Stoehr, Felix and
                  Saini, Kamaljeet},
  month        = mar,
  year         = 2022,
  doi          = {10.5281/zenodo.7764458},
  url          = {https://doi.org/10.5281/zenodo.7764458},
}

@manual{cortes_2025_14933753,
  title        = {ALMA Cycle 12 Technical Handbook},
  author       = {Cortes, Paulo and
                  Carpenter, John and
                  Kameno, Seiji and
                  Loomis, Ryan and
                  Vila Vilaro, Baltasar and
                  Immer, Katharina and
                  Vlahakis, Catherine and
                  Law, James and
                  Stoehr, Felix and
                  Saini, Kamaljeet and
                  Hales, Antonio and
                  Kneissl, Ruediger},
  month        = mar,
  year         = 2025,
  doi          = {10.5281/zenodo.14933753},
  url          = {https://doi.org/10.5281/zenodo.14933753},
}

@ARTICLE{2021NatAs...5..631V,
       author = {{Villanueva}, G.~L. and {Cordiner}, M. and {Irwin}, P.~G.~J. and {de Pater}, I. and {Butler}, B. and {Gurwell}, M. and {Milam}, S.~N. and {Nixon}, C.~A. and {Luszcz-Cook}, S.~H. and {Wilson}, C.~F. and {Kofman}, V. and {Liuzzi}, G. and {Faggi}, S. and {Fauchez}, T.~J. and {Lippi}, M. and {Cosentino}, R. and {Thelen}, A.~E. and {Moullet}, A. and {Hartogh}, P. and {Molter}, E.~M. and {Charnley}, S. and {Arney}, G.~N. and {Mandell}, A.~M. and {Biver}, N. and {Vandaele}, A.~C. and {de Kleer}, K.~R. and {Kopparapu}, R.},
        title = "{No evidence of phosphine in the atmosphere of Venus from independent analyses}",
      journal = {Nature Astronomy},
         year = 2021,
        month = jan,
       volume = {5},
        pages = {631-635},
          doi = {10.1038/s41550-021-01422-z},
archivePrefix = {arXiv},
       eprint = {2010.14305},
 primaryClass = {astro-ph.EP},
       adsurl = {https://ui.adsabs.harvard.edu/abs/2021NatAs...5..631V}
}

@ARTICLE{2018AJ....155..182L,
       author = {{Loomis}, Ryan A. and {{\"O}berg}, Karin I. and {Andrews}, Sean M. and {Walsh}, Catherine and {Czekala}, Ian and {Huang}, Jane and {Rosenfeld}, Katherine A.},
        title = "{Detecting Weak Spectral Lines in Interferometric Data through Matched Filtering}",
      journal = {\aj},
         year = 2018,
        month = apr,
       volume = {155},
       number = {4},
          eid = {182},
        pages = {182},
          doi = {10.3847/1538-3881/aab604},
archivePrefix = {arXiv},
       eprint = {1803.04987},
 primaryClass = {astro-ph.IM},
       adsurl = {https://ui.adsabs.harvard.edu/abs/2018AJ....155..182L}
}

@article{DBLP:journals/corr/ChenG16,
  author       = {Tianqi Chen and
                  Carlos Guestrin},
  title        = {XGBoost: {A} Scalable Tree Boosting System},
  journal      = {CoRR},
  volume       = {abs/1603.02754},
  year         = {2016},
  url          = {http://arxiv.org/abs/1603.02754},
  eprinttype   = {arXiv},
  eprint       = {1603.02754},
  bibsource    = {dblp computer science bibliography, https://dblp.org}
}

@ARTICLE{2023PASP..135g4501H,
       author = {{Hunter}, Todd R. and {Indebetouw}, Remy and {Brogan}, Crystal L. and {Berry}, Kristin and {Chang}, Chin-Shin and {Francke}, Harold and {Geers}, Vincent C. and {G{\'o}mez}, Laura and {Hibbard}, John E. and {Humphreys}, Elizabeth M. and {Kent}, Brian R. and {Kepley}, Amanda A. and {Kunneriath}, Devaky and {Lipnicky}, Andrew and {Loomis}, Ryan A. and {Mason}, Brian S. and {Masters}, Joseph S. and {Maud}, Luke T. and {Muders}, Dirk and {Sabater}, Jose and {Sugimoto}, Kanako and {Sz{\H{u}}cs}, L{\'a}szl{\'o} and {Vasiliev}, Eugene and {Videla}, Liza and {Villard}, Eric and {Williams}, Stewart J. and {Xue}, Rui and {Yoon}, Ilsang},
        title = "{The ALMA Interferometric Pipeline Heuristics}",
      journal = {\pasp},
         year = 2023,
        month = jul,
       volume = {135},
       number = {1049},
          eid = {074501},
        pages = {074501},
          doi = {10.1088/1538-3873/ace216},
archivePrefix = {arXiv},
       eprint = {2306.07420},
 primaryClass = {astro-ph.IM},
       adsurl = {https://ui.adsabs.harvard.edu/abs/2023PASP..135g4501H}
}

@article{scikit-learn,
  title={Scikit-learn: Machine Learning in {P}ython},
  author={Pedregosa, F. and Varoquaux, G. and Gramfort, A. and Michel, V.
          and Thirion, B. and Grisel, O. and Blondel, M. and Prettenhofer, P.
          and Weiss, R. and Dubourg, V. and Vanderplas, J. and Passos, A. and
          Cournapeau, D. and Brucher, M. and Perrot, M. and Duchesnay, E.},
  journal={Journal of Machine Learning Research},
  volume={12},
  pages={2825--2830},
  year={2011}
}

@Manual{xgboost,
  title = {xgboost: Extreme Gradient Boosting},
  author = {Tianqi Chen and Tong He and Michael Benesty and Vadim Khotilovich and Yuan Tang and Hyunsu Cho and Kailong Chen and Rory Mitchell and Ignacio Cano and Tianyi Zhou and Mu Li and Junyuan Xie and Min Lin and Yifeng Geng and Yutian Li and Jiaming Yuan and David Cortes},
  year = {2025},
  note = {R package version 3.1.1.1},
  url = {https://github.com/dmlc/xgboost},
}

@ARTICLE{2022ApJ...938..151E,
       author = {{Ewall-Wice}, Aaron and {Dillon}, Joshua S. and {Gehlot}, Bharat and {Parsons}, Aaron and {Cox}, Tyler and {Jacobs}, Daniel C.},
        title = "{Precision Calibration of Radio Interferometers for 21 cm Cosmology with No Redundancy and Little Knowledge of Antenna Beams and the Radio Sky}",
      journal = {\apj},
         year = 2022,
        month = oct,
       volume = {938},
       number = {2},
          eid = {151},
        pages = {151},
          doi = {10.3847/1538-4357/ac87b3},
archivePrefix = {arXiv},
       eprint = {2110.11994},
 primaryClass = {astro-ph.CO},
       adsurl = {https://ui.adsabs.harvard.edu/abs/2022ApJ...938..151E}
}

@ARTICLE{2025A&A...703A.196A,
       author = {{{\'A}lvarez-Ortega}, D. and {Casadio}, C. and {P{\"o}tzl}, F.~M. and {Kumar}, A. and {Janssen}, M.},
        title = "{VIPCALs: A fully automated calibration pipeline for very long baseline interferometry data}",
      journal = {\aap},
         year = 2025,
        month = nov,
       volume = {703},
          eid = {A196},
        pages = {A196},
          doi = {10.1051/0004-6361/202556838},
archivePrefix = {arXiv},
       eprint = {2508.13282},
 primaryClass = {astro-ph.IM},
       adsurl = {https://ui.adsabs.harvard.edu/abs/2025A&A...703A.196A}
}

@ARTICLE{2001ITAP...49.1683P,
       author = {{Pardo}, Juan R. and {Cernicharo}, Jos{\'e} and {Serabyn}, Eugene},
        title = "{Atmospheric Transmission at Microwaves (ATM): An Improved Model for Millimeter/Submillimeter Applications}",
      journal = {IEEE Transactions on Antennas and Propagation},
         year = 2001,
        month = jan,
       volume = {49},
       number = {12},
        pages = {1683},
          doi = {10.1109/8.982447},
       adsurl = {https://ui.adsabs.harvard.edu/abs/2001ITAP...49.1683P}
}

@ARTICLE{2010MNRAS.406L..60N,
       author = {{Nita}, G.~M. and {Gary}, D.~E.},
        title = "{The generalized spectral kurtosis estimator}",
      journal = {\mnras},
         year = 2010,
        month = jul,
       volume = {406},
       number = {1},
        pages = {L60-L64},
          doi = {10.1111/j.1745-3933.2010.00882.x},
archivePrefix = {arXiv},
       eprint = {1005.4371},
 primaryClass = {astro-ph.IM},
       adsurl = {https://ui.adsabs.harvard.edu/abs/2010MNRAS.406L..60N}
}

@inproceedings{GarciaYus_2020,
    author = {Jorge F. Garc{\'i}a Yus and Bill Dent and Drew Brisbin and Chin-Shin Chang and Laura G{\'o}mez and Theodoros Nakos},
    title = {{Towards the processing, review, and delivery of 80% of the ALMA data by the Joint ALMA Observatory (JAO)}},
    volume = {11449},
    booktitle = {Observatory Operations: Strategies, Processes, and Systems VIII},
    editor = {David S. Adler and Robert L. Seaman and Chris R. Benn},
    organization = {International Society for Optics and Photonics},
    publisher = {SPIE},
    pages = {114491U},
    year = {2020},
    doi = {10.1117/12.2560719},
    URL = {https://doi.org/10.1117/12.2560719}
}

@ARTICLE{2007A&A...462..801E,
       author = {{Escoffier}, R.~P. and {Comoretto}, G. and {Webber}, J.~C. and {Baudry}, A. and {Broadwell}, C.~M. and {Greenberg}, J.~H. and {Treacy}, R.~R. and {Cais}, P. and {Quertier}, B. and {Camino}, P. and {Bos}, A. and {Gunst}, A.~W.},
        title = "{The ALMA correlator}",
      journal = {\aap},
         year = 2007,
        month = feb,
       volume = {462},
       number = {2},
        pages = {801-810},
          doi = {10.1051/0004-6361:20054519},
       adsurl = {https://ui.adsabs.harvard.edu/abs/2007A&A...462..801E}
}

@ARTICLE{2010MNRAS.405..155O,
       author = {{Offringa}, A.~R. and {de Bruyn}, A.~G. and {Biehl}, M. and {Zaroubi}, S. and {Bernardi}, G. and {Pandey}, V.~N.},
        title = "{Post-correlation radio frequency interference classification methods}",
      journal = {\mnras},
         year = 2010,
        month = jun,
       volume = {405},
       number = {1},
        pages = {155-167},
          doi = {10.1111/j.1365-2966.2010.16471.x},
archivePrefix = {arXiv},
       eprint = {1002.1957},
 primaryClass = {astro-ph.IM},
       adsurl = {https://ui.adsabs.harvard.edu/abs/2010MNRAS.405..155O}
}

@inproceedings{Nakos_2020,
    author = {Theodoros Nakos and Harold Francke and Kouichiro Nakanishi and Dirk Petry and Thomas Stanke and Catarina Ubach and Luciano Cerrigone and Erica Keller and Alfonso Trejo and Junko Ueda},
    title = {{Improving ALMA’s data processing efficiency using a holistic approach}},
    volume = {11449},
    booktitle = {Observatory Operations: Strategies, Processes, and Systems VIII},
    editor = {David S. Adler and Robert L. Seaman and Chris R. Benn},
    organization = {International Society for Optics and Photonics},
    publisher = {SPIE},
    pages = {114491T},
    year = {2020},
    doi = {10.1117/12.2559324},
    URL = {https://doi.org/10.1117/12.2559324}
}

@ARTICLE{6846297,
  author={Gavish, Matan and Donoho, David L.},
  journal={IEEE Transactions on Information Theory}, 
  title={The Optimal Hard Threshold for Singular Values is  $4/\sqrt {3}$ }, 
  year={2014},
  volume={60},
  number={8},
  pages={5040-5053},
  doi={10.1109/TIT.2014.2323359}}

@ARTICLE{2019ApJ...871..112X,
       author = {{Xue}, Ci and {Remijan}, Anthony J. and {Burkhardt}, Andrew M. and
         {Herbst}, Eric},
        title = "{ALMA Observations of the Spatial Distribution of Three C$_{2}$H$_{4}$O$_{2}$ Isomers toward Sgr B2(N)}",
      journal = {\apj},
         year = "2019",
        month = "Jan",
       volume = {871},
       number = {1},
          eid = {112},
        pages = {112},
          doi = {10.3847/1538-4357/aaf738},
archivePrefix = {arXiv},
       eprint = {1812.07091},
 primaryClass = {astro-ph.GA},
       adsurl = {https://ui.adsabs.harvard.edu/abs/2019ApJ...871..112X}
}

@manual{remijan_2019_4511522,
  title        = {ALMA Cycle 7 Technical Handbook},
  author       = {Remijan, Anthony and
                  Biggs, Andy and
                  Cortes, Paulo and
                  Dent, Bill and
                  Di Francesco, James and
                  Fomalont, Ed and
                  Hales, Antonio and
                  Kameno, Seiji and
                  Mason, Brian and
                  Philips, N. and
                  Saini, Kamaljeet and
                  Stoehr, Felix and
                  Vila Vilaro, Baltasar and
                  Villard, Eric},
  month        = jun,
  year         = 2019,
  doi          = {10.5281/zenodo.4511522},
  url          = {https://doi.org/10.5281/zenodo.4511522},
}

@ARTICLE{2022RemS...14.4672P,
       author = {{Perez-Portero}, Adrian and {Querol}, Jorge and {Camps}, Adriano and {Martin-Neira}, Manuel and {Suess}, Martin and {Ramirez}, Juan Ignacio and {Zurita}, Alberto and {Closa}, Josep and {Oliva}, Roger and {Onrubia}, Raul},
        title = "{RFI Detection and Mitigation for Advanced Correlators in Interferometric Radiometers}",
      journal = {Remote Sensing},
         year = 2022,
        month = sep,
       volume = {14},
       number = {18},
        pages = {4672},
          doi = {10.3390/rs14184672},
       adsurl = {https://ui.adsabs.harvard.edu/abs/2022RemS...14.4672P}
}

@manual{almapipeline2025,
  title        = {ALMA Science Pipeline User's Guide for Release 2025.1.0.37},
  author       = {{ALMA/CASA Pipeline Team}},
  organization = {Joint ALMA Observatory},
  year         = {2025},
  url          = {https://almascience.nrao.edu/processing/documents-and-tools/cycle12/alma_pipeline_users_guide_2025},
  note         = {Accessed: 2026-08-07; Packaged with CASA 6.6.1}
}

@dataset{xue_2026_21972118,
  author       = {Xue, Ci and
                  Mason, Brian and
                  Ashton, Tristan},
  title        = {Training dataset of the selected ALMA Cycle 9
                   bandpass calibration solutions
                  },
  month        = aug,
  year         = 2026,
  publisher    = {Zenodo},
  doi          = {10.5281/zenodo.21972118},
  url          = {https://doi.org/10.5281/zenodo.21972118},
}

@ARTICLE{rakib2026efficientregressionmodelsscan,
      title={Efficient Regression Models for Scan Statistics}, 
      author={Gazi Abdur Rakib and Tristan Ashton and Ryan A. Loomis and Brian S. Mason and Eric J. Murphy and Ci Xue and Jeff M. Phillips},
      year={2026},
      eprint={2608.22201},
      archivePrefix={arXiv},
      primaryClass={stat.ME},
      url={https://arxiv.org/abs/2608.22201}, 
}

@article{Kulldorff1997,
    author = {Martin Kulldorff},
    title = {A spatial scan statistic},
    journal = {Communications in Statistics - Theory and Methods},
    volume = {26},
    number = {6},
    pages = {1481--1496},
    year = {1997},
    publisher = {Taylor \& Francis},
    doi = {10.1080/03610929708831995},
    URL = {https://doi.org/10.1080/03610929708831995},
    eprint = {https://doi.org/10.1080/03610929708831995}
}

@BOOK{2024hss..book.....G,
       author = {{Glaz}, Joseph and {Koutras}, Markos V.},
        title = "{Handbook of Scan Statistics}",
         year = 2024,
    publisher = {Springer},
          doi = {10.1007/978-1-4614-8033-4},
    isbn = {978-1-4614-8033-4},
}

@INPROCEEDINGS{2024pcaw.confE..12B,
       author = {{Brogan}, Crystal},
        title = "{The Advanced Technology ALMA Correlator}",
    booktitle = {The Promises and Challenges of the ALMA Wideband Sensitivity Upgrade},
         year = 2024,
        month = sep,
          eid = {12},
        pages = {12},
          doi = {10.5281/zenodo.13681462},
       adsurl = {https://ui.adsabs.harvard.edu/abs/2024pcaw.confE..12B}
}

@INPROCEEDINGS{2024stt..conf....1D,
       author = {{Donovan Meyer}, Jennifer and {Loomis}, Ryan A. and {Brogan}, Crystal and {Hunter}, Todd},
        title = "{Upgrading the Future of ALMA: the Wideband Sensitivity Upgrade}",
    booktitle = {33rd IEEE International Symposium on Space THz Technology (ISSTT 2024)},
         year = 2024,
        month = apr,
        pages = {1},
       adsurl = {https://ui.adsabs.harvard.edu/abs/2024stt..conf....1D}
}
\bibliographystyle{aasjournal}

\appendix
\section{Misclassified Solutions \label{apx:misclassification}}
Figures~\ref{fig:fn} and \ref{fig:fp} show misclassified bandpass calibration solutions over the five outer test folds of the nested cross-validation. Each panel shows a pair of amplitude solutions for the two orthogonal polarizations, identified by the EB ID, antenna name, and spectral window index in the lower right.

\vspace{-0.5em}
\begin{figure}[h!]
    \centering
    \includegraphics[width=0.81\linewidth]{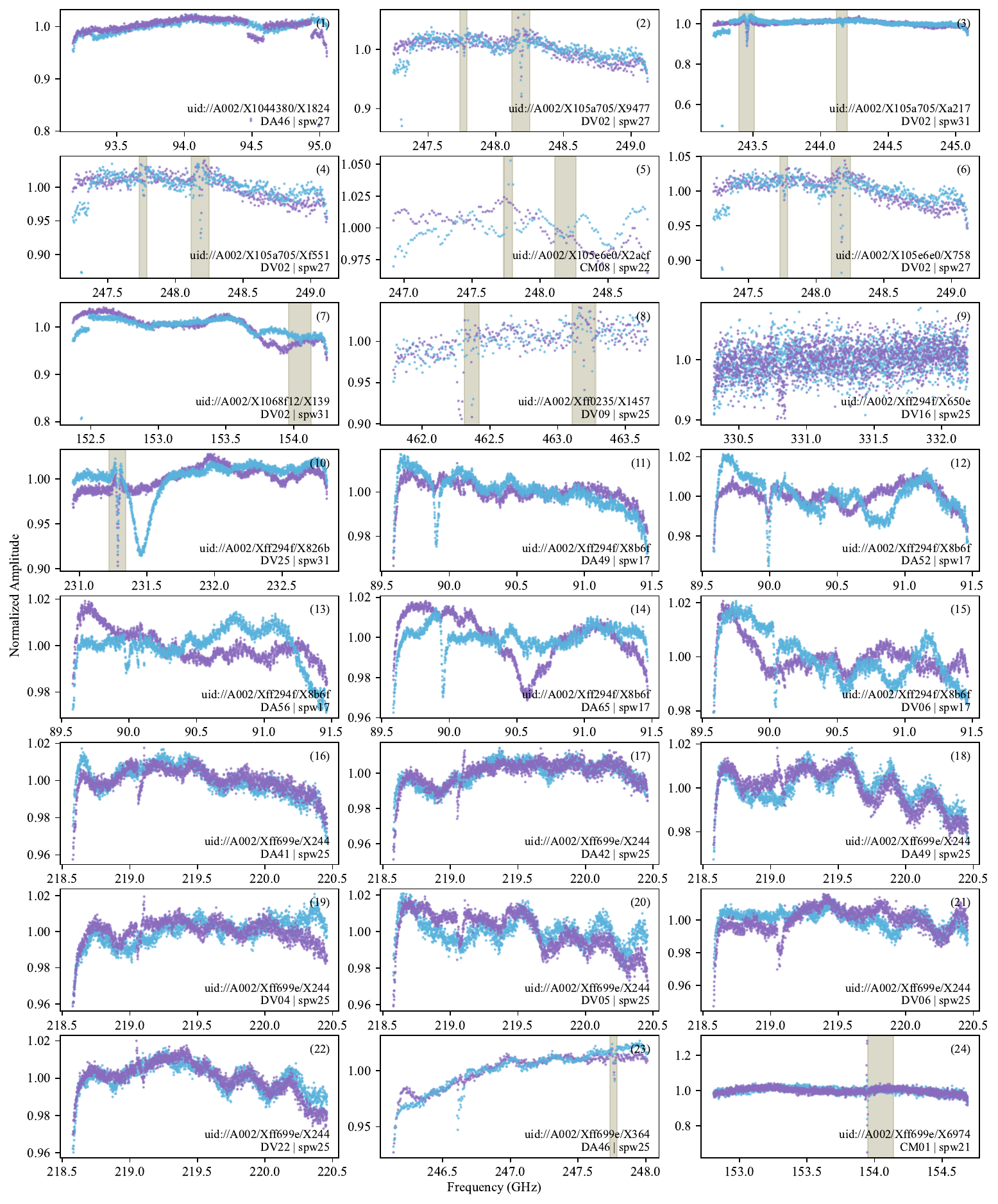}
    \caption{\footnotesize False negative classifications in the selected Cycle 9 training dataset. 24 pairs of amplitude solutions (polarization X in blue and Y in purple) were predicted as nominal solutions (\texttt{False}), whereas QA2 labeled them anomalous (\texttt{True}). Brown shaded regions mark atmospheric interference, where present.}
    \label{fig:fn}
\end{figure}

\begin{figure}
    \centering
    \includegraphics[width=0.95\linewidth]{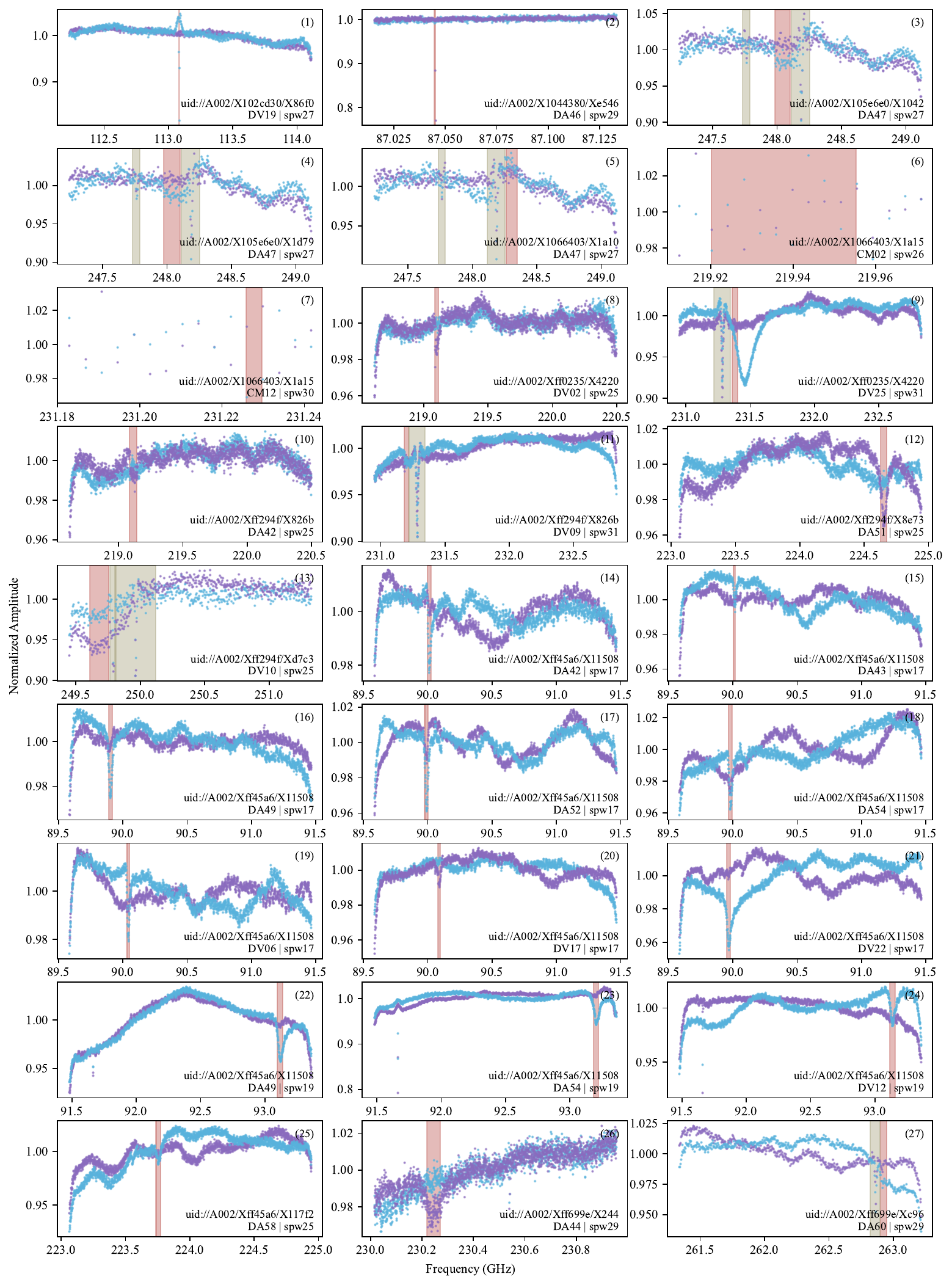}
    \caption{\footnotesize False positive classifications in the selected Cycle 9 training dataset. 27 pairs of amplitude solutions (polarization X in blue and Y in purple) were predicted as anomalous (\texttt{True}), whereas the QA2 labels are \texttt{False}. Brown shaded regions mark atmospheric interference, where present. Red shaded regions mark anomalous intervals identified by the scan-statistic masked scoring.}
    \label{fig:fp}
\end{figure}

\end{document}